\documentclass[a4paper,12pt]{article}
\usepackage{color}
\usepackage{amsfonts}
\usepackage{amsmath}
\usepackage{graphicx}
\usepackage{epsfig}
\usepackage{braket}
\newcommand{\new}{\textcolor{black}}

\def\SU{\text{SU}}

\def\be{\begin{equation}}
\def\ee{\end{equation}}
\def\beq{\begin{equation}}
\def\eeq{\end{equation}}
\def\bea{\begin{eqnarray}}
\def\eea{\end{eqnarray}}

\def\<{\left\langle}
\def\>{\right\rangle}

\allowdisplaybreaks[2]
\begin{document}
\bibliographystyle{OurBibTeX}
\begin{titlepage}
 \vspace*{-15mm}
\begin{flushright}
\end{flushright}
\vspace*{5mm}
\begin{center}
{ \sffamily \LARGE $A_4$ Family Symmetry from $SU(5)$ SUSY GUTs in
6d}
\\[8mm]
T.~J.~Burrows\footnote{E-mail: \texttt{tjb54@soton.ac.uk}} and
S.~F.~King\footnote{E-mail: \texttt{king@soton.ac.uk}}
\\[3mm]
{\small\it
School of Physics and Astronomy,
University of Southampton,\\
Southampton, SO17 1BJ, U.K.
}\\[1mm]
\end{center}
\vspace*{0.75cm}
\begin{abstract}
\noindent We propose a model in which $A_4$ Family Symmetry arises
dynamically from a six dimensional orbifold $SU(5)$ Supersymmetric
Grand Unified Theory. The $SU(5)$ is broken to the Standard Model
gauge group by a particular orbifold compactification leading to
$A_4$ Family Symmetry, low energy Supersymmetry and Higgs
doublet-triplet splitting. The resulting four dimensional
effective superpotential leads to a realistic description of quark
and lepton masses and mixing angles including tri-bimaximal
neutrino mixing and an inter-family mass hierarchy
provided by a Froggatt-Nielsen mechanism. This model is the first
which combines the idea of orbifold GUTs with $A_4$
family symmetry resulting from the orbifolding.
\end{abstract}
\end{titlepage}
\newpage
\setcounter{footnote}{0}

\section{Introduction}

It is well known that the solar and atmospheric data are
consistent with so-called tri-bimaximal (TB) mixing
\cite{HPS},
\begin{eqnarray}
U_{TB} =
\left( \begin{array}{rrr}
-\frac{2}{\sqrt{6}}   & \frac{1}{\sqrt{3}} & 0 \\
\frac{1}{\sqrt{6}}  & \frac{1}{\sqrt{3}} & \frac{1}{\sqrt{2}} \\
\frac{1}{\sqrt{6}}  & \frac{1}{\sqrt{3}} & -\frac{1}{\sqrt{2}}
\end{array}
\right).
\label{MNS0}
\end{eqnarray}
The ansatz of TB lepton mixing matrix is interesting due to its
symmetry properties which seem to call for a possibly discrete
non-Abelian Family Symmetry in nature \cite{Harrison:2003aw}.
There has been a considerable amount of theoretical work in which
the observed TB neutrino flavour symmetry may be related to some
Family Symmetry
\cite{Chen:2009um,King:2005bj,deMedeirosVarzielas:2005ax,deMedeirosVarzielas:2006fc,King:2006me,
Altarelli:2006ri,Frampton:2004ud,King:2009mk,Lam:2008sh,Grimus:2009pg}.
These models may be classified according to the way that TB mixing
is achieved, namely either directly or indirectly
\cite{King:2009ap}. The direct models are based on $A_4$ or $S_4$,
or a larger group that contains these groups as a subgroup, and in
these models some of the generators of the Family Symmetry survive
to form at least part of the neutrino flavour symmetry. In the
indirect models, typically based on $\Delta(3n^2)$ or
$\Delta(6n^2)$, none of the generators of the Family Symmetry
appear in the neutrino flavour symmetry \cite{King:2009ap}.

The most ambitious models combine Family Symmetry with grand
unified theories (GUTs). The minimal Family Symmetry which
contains triplet representations and can lead to TB mixing via the
direct model approach is $A_4$. The minimal simple GUT group is
$SU(5)$. A direct model has been proposed which combines $A_4$
Family Symmetry with $SU(5)$ Supersymmetric (SUSY) GUTs
\cite{Altarelli:2008bg}. This model was formulated in five
dimensions (5d), in part to address the doublet-triplet splitting
problem of GUTs, and in part to allow a viable description of the
charged fermion mass hierarchies, by placing the lightest two
tenplets $T_1$, $T_2$ in the bulk, while the pentaplets $F$ and
$T_3$ are on the brane. An additional $U(1)$ Family Symmetry is
also assumed in order to yield hierarchies between different
families via the Froggatt-Nielsen mechanism
\cite{Froggatt:1978nt}.

In the approach in \cite{Altarelli:2008bg} the $A_4$ is simply
assumed to exist in the 5d theory. However it has been shown how
an $A_4$ Family Symmetry could have a dynamical origin as a result
of the compactification of a 6d theory down to 4d
\cite{Altarelli:2006kg}. Similar considerations have been applied
to other discrete family symmetries \cite{Adulpravitchai:2009id},
and the connection to string theory of these and other orbifold
compactifications has been discussed in \cite{Kobayashi:2006wq}.
According to \cite{Altarelli:2006kg}, the $A_4$ appears as a
symmetry of the orbifold fixed points on which 4d branes, which
accommodate the matter fields, reside, while the flavons which
break $A_4$ are in the bulk. The formulation of a theory in 6d is
also closer in spirit to string theories which are formulated in
10d where such theories are often compactified in terms of three
complex compact dimensions. The 6d theory here will involve one
complex compact dimension $z$.

The purpose of this paper is to formulate a realistic direct model
in which an $A_4$ Family Symmetry arises dynamically from an
$SU(5)$ SUSY GUT in 6d. The $A_4$ Family Symmetry emerges as a
result of the compactification of the extra complex compact
dimension $z$, assuming a particular orbifolding. $SO(10)$ in 6d
has been considered in \cite{Asaka:2001eh}, with the extra
dimensions compactified on a rectangular torus. In order to
realize an $A_4$ Family Symmetry upon compactification, we shall
generalise the formalism of 6d GUTs in \cite{Asaka:2001eh} to the
case of compactification on a twisted torus. Then, starting from
an $SU(5)$ SUSY GUT in 6d, we shall show how the $A_4$ Family
Symmetry can result from the symmetry of the orbifold fixed points
after compactification, assuming a particular twist angle $\theta
= 60^\circ$ and a particular orbifold
$\mathbb{T}^2/(\mathbb{Z}_2\times\mathbb{Z}_2^{\mathrm{SM}})$.
Unlike the model in \cite{Altarelli:2008bg}, the resulting model
has all three tenplets $T_i$, as well as the pentaplet $F$,
located on the 3-branes at the fixed points. However, as in
\cite{Altarelli:2008bg}, we shall assume an additional $U(1)$
Froggatt-Nielsen Family Symmetry to account for inter-family mass
hierarchies. We emphasise that this model is the first
which combines the idea of orbifold GUTs with $A_4$
family symmetry resulting from the orbifolding.

The layout of the remainder of the paper is as follows. In Section
2 we generalize the formulation of 6d GUTs (usually compactified
on a rectangular torus) to the general case of compactification on
a twisted torus with a general twist angle $\theta$. Then we show
how compactification of the $SU(5)$ SUSY GUT in 6d on an orbifold
$\mathbb{T}^2/(\mathbb{Z}_2\times\mathbb{Z}_2^{\mathrm{SM}})$
leads to an effective 4d theory with $\mathcal{N}=1$ SUSY
preserved but the $SU(5)$ GUT broken to the Standard Model (SM)
gauge group. We also show how Higgs doublet-triplet splitting
emerges if the Higgs fields are in the bulk. In Section \new{\ref{modelsection}}
we present the $SU(5)$ SUSY GUT model in 6d in which the $A_4$
Family Symmetry emerges after the above compactification. We
specify the superfield content and symmetries of the model and
provide a dictionary for the realization of the 4d effective
superpotential in terms of the 6d $A_4$ invariants. From the
effective 4d superpotential we show how a successful pattern of
quark and lepton masses and mixing, including tri-bimaximal
neutrino mixing, can emerge. In Section \new{\ref{alignmentsection}} we comment on the vacuum
alignment and subleading corrections expected in the model.
Section \new{\ref{conclusion}} concludes the paper. In order to make the paper
self-contained we include an Appendix on the $A_4$ group and it's
representations. We also include another Appendix which
summarizes how $A_4$ family symmetry can arise from the orbifold
discussed in this paper.

\section{$SU(5)$ GUTs in six dimensions on a twisted torus}
\subsection{The gauge sector of SUSY $SU(5)$ in 6d}
We are considering a $\mathcal{N}=1$ supersymmetric Yang-Mills
theory in 6 dimensions, the Lagrangian reads,

\begin{equation}
\mathcal{L}^{YM}_{6d}=
\mathrm{Tr}(-\frac{1}{2}V_{MN}V^{MN}+i\overline{\Lambda}\Gamma^MD_{M}\Lambda),
\end{equation}
where $V_{M}= t^aV^{a}_{M}$ and $\Lambda = t^a\Lambda^a$, here
$t^a$ are the generators of SU(5). $D_M\Lambda=
\partial_m\Lambda-ig[V_M,\Lambda]$ and $V_{MN}= [D_M,D_N]/(ig)$.
The $\Gamma$ matrices are given by:
\begin{align}
\Gamma^\mu =
 \begin{pmatrix}
   \gamma^\mu & 0 \\
   0 & \gamma^\mu
 \end{pmatrix},\qquad
 \Gamma^5 =
 \begin{pmatrix}
   0 & i\gamma^5  \\
  i\gamma^5 & 0
 \end{pmatrix},\qquad
\Gamma^6 =
 \begin{pmatrix}
   0 & \gamma^5 \\
  -\gamma^5 & 0
 \end{pmatrix}
\end{align}
with $\gamma^5 = I$ and $\{ \Gamma_M, \Gamma_N
\}=2\eta_{MN}\mathbf{1}_{(8\times 8)},\eta_{MN}=\mathrm{diag}(1,-1,-1,-1,-1,-1).$ The gaugino
$\Lambda$ is composed of two Weyl fermions of opposite chirality
in 4d,
\begin{eqnarray}
 \Lambda=(\lambda_1, -i\lambda_2), &\gamma_5\lambda_1=-\lambda_1,& \gamma_5\lambda_2=\lambda_2.
\end{eqnarray}
Overall the gaugino has negative 6d chirality
$\Gamma_7\Lambda=-\Lambda$, where
$\Gamma_7=\mathrm{diag}(\gamma_5,-\gamma_5).$

\subsection{Compactification on a twisted torus}

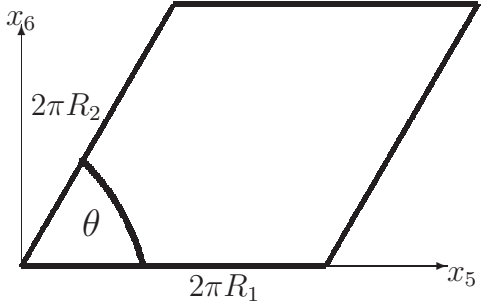
\begin{figure}
\ifx\JPicScale\undefined\def\JPicScale{.4}\fi \unitlength
\JPicScale mm
\begin{picture}(160,95)(0,-10)
\linethickness{0.6mm} \put(10,0){\line(1,0){100}}
\linethickness{0.6mm} \put(60,87){\line(1,0){100}}
\linethickness{0.6mm}
\multiput(10,0)(0.12,0.208){417}{\line(0,1){0.2}}
\linethickness{0.6mm}
\multiput(110,0)(0.12,0.208){417}{\line(0,1){0.21}}
\linethickness{0.1mm} \put(10,0){\vector(1,0){140}}
\put(10,0){\vector(0,1){80}}
\put(150,-5){$x_5$} \put(5,80){$x_6$} \put(65,-10){$2\pi R_1$}
\put(13,50){$2\pi R_2$} \large \put(30,10){$\theta$}
\linethickness{0.6mm} \qbezier(30,35)(45,20)(50,0)
\end{picture}
\caption{\small The twisted torus, $R_1$ and $R_2$ are the radii
and $\theta$ is the twist angle (later we shall specify
$\theta=\pi/3$ and $R_1=2R_2$). } \label{twistedtorus}
\end{figure}

We compactify the two extra dimensions on a twisted torus $\mathbb{T}^2$
so that the theory lives on $M=\mathcal{R}^4\times\mathbb{T}^2$. The torus is defined by:

\begin{eqnarray}
\label{twistedsymmetry}
(x_5,x_6)&\rightarrow &(x_5 +2\pi R_1,x_6) \\
(x_5,x_6) &\rightarrow &(x_5+2\pi R_2\cos\theta, x_6 + 2\pi R_2\sin\theta).
\end{eqnarray}
We can expand the $SU(5)$ gauge multiplet fields $\Phi=(V_M, \Lambda)$ using the mode
expansion:
\begin{equation}
\Phi(x,x_5,x_6)=\frac{1}{2\pi\sqrt{R_1R_2\sin\theta}}\sum_{m,n}\Phi^{(m,n)}(x)\exp\left\lbrace
i\left(\frac{m}{R_1}\{x_5-\frac{x_6}{\tan\theta}\}+\frac{nx_6}{R_2\sin\theta}\right)\right\rbrace ,
\end{equation}
where $R_1$ and $R_2$ are the two radii of the torus and $\theta$ is the angle of twist as shown in
Fig.\ref{twistedtorus}. The vector field is Hermitian so the
coefficients satisfy the relation
$V_M^{(-m,-n)}=V_M^{(m,n)\dagger}$. To obtain the 4d effective
Lagrangian we integrate over the extra dimensions. Note that we
are only including terms below $\mathcal{O}(1/R)$ so there are
only bilinear terms in the 4d Lagrangian. We make a convenient
choice of variables for the 4d scalars:

\begin{eqnarray}
 \Pi_1^{(m,n)}(x) = \frac{i}{M(m,n)}\left(\frac{m}{R_1}V_5^{(m,n)}(x)+\left(\frac{m}{R_1\tan\theta}-\frac{n}{R_2\sin\theta}\right)V_6^{(m,n)}(x)\right)\\
\Pi_2^{(m,n)}(x) =
\frac{i}{M(m,n)}\left(-\left(\frac{m}{R_1\tan\theta}-\frac{n}{R_2\sin\theta}\right)V_5^{(m,n)}(x)+\frac{m}{R_1}V_6^{(m,n)}(x)\right)
\end{eqnarray}
where $M(m,n)=\frac{1}{\sin\theta}\sqrt{\left(\frac{m}{R_1}\right)^2
+\left(\frac{n}{R_2}\right)^2-\frac{2mn\cos\theta}{R_1R_2}}$. The
4d Lagrangian for the gauge and scalar fields is then given by:

\begin{align}
 \mathcal{L}^{(1)}_{4d} =& \sum_{m,n}  \mathrm{Tr}(-\frac{1}{2}\tilde{V}^{(m,n)\dagger}_{\mu\nu}\tilde{V}^{(m,n)\mu\nu} +M(m,n)^2 V^{(m,n)\dagger}_{\mu}V^{(m,n)\mu}\nonumber \\
  & + \partial_\mu\Pi_2^{(m,n)\dagger}\partial^\mu\Pi_2^{(m,n)\dagger} +M(m,n)^2\Pi_2^{(m,n)\dagger}\Pi_2^{(m,n)}\nonumber\\
  & + \partial_\mu\Pi_1^{(m,n)\dagger}\partial^\mu\Pi_1^{(m,n)}\nonumber\\
  & - M(m,n)(V^{(m,n)\dagger}_\mu\partial^\mu\Pi_1^{(m,n)}+\partial^\mu\Pi_1^{(m,n)\dagger}V_\mu^{(m,n)} ))
\end{align}
where $\tilde{V}^{(m,n)}_{\mu\nu}=\partial_\mu
V_\nu^{(m,n)}-\partial_\nu V_\mu^{(m,n)}$. The gaugino part of the
Lagrangian integrates to

\begin{align}
 \mathcal{L}^{(2)}_4  = &\sum_{m,n} \mathrm{Tr}( i\overline{\lambda}^{(m,n)}_1\gamma^\mu\partial_\mu\lambda_1^{(m,n)} + i\overline{\lambda}^{(m,n)}_2\gamma^\mu\partial_\mu\lambda_2^{(m,n)}\nonumber\\
 &
-\left(\frac{m}{R_1}-i\left(\frac{n}{R_2\sin\theta}-\frac{m}{R_1\tan\theta}\right)\right)\overline{\lambda}^{(m,n)}_1\lambda_2^{(m,n)}
+ c.c.).
\end{align}
This is the kinetic term for a dirac fermion $\lambda_D =
(\lambda_1, \lambda_2)$ with a mass $M(m,n)$. In total there is
the vector $V_\mu^{(m,n)}$, scalars $\Pi_{1,2}^{(m,n)}$ and
$\lambda_D$ forming a massive $\mathcal{N}=1$ vector multiplet in
4d. However when we look at the massless sector of the theory we
have unwanted $\mathcal{N}=2$ symmetry which can be removed by
orbifolding, as we now discuss.

\subsection{Compactification on the orbifold $\mathbb{T}^2/\mathbb{Z}_2$}
\label{orbifoldsection}

Instead of compactifying on the torus we can compactify on the
orbifold $\mathbb{T}^2/\mathbb{Z}_2$ where we assign parities
under the reflection $(x_5,x_6)\rightarrow(-x_5,-x_6)$ to the
vectors and scalars:
\begin{eqnarray}
 PV_\mu(x,-x_5,-x_6)P^{-1} & = & +V_\mu(x,x_5,x_6)\\
PV_{5,6}(x,-x_5,-x_6)P^{-1} & = & -V_{5,6}(x,x_5,x_6),
\end{eqnarray}
where we chose $P=I$, so for the Fourier modes we find:
\begin{eqnarray}
 V_\mu^{(-m,-n)}=+V_\mu^{(m,n)}&=&+V_\mu^{(m,n)\dagger},\\
V_{5,6}^{(-m,-n)}=-V_{5,6}^{(m,n)}&=&+V_{5,6}^{(m,n)\dagger}.
\end{eqnarray}
This eliminates the scalar zero modes, also the number of massive
modes is halved. Because the derivatives $\partial_{5,6}$ are odd
under the reflection the two Weyl fermions must have opposite
parities:
\begin{eqnarray}
 P\lambda_1(x,-x_5,-x_6)P^{-1}&=&+\lambda_1(x,x_5,x_6)\\
P\lambda_2(x,-x_5,-x_6)P^{-1}&=&-\lambda_2(x,x_5,x_6)
\end{eqnarray}
$(V_\mu,\lambda_1)$ and $(V_{5,6},\lambda_2)$ form vector and
chiral multiplets respectively, only the vector multiplets have
zero modes. The orbifolding has thus broken the extended
$\mathcal{N}=2$ SUSY in 4d down to $\mathcal{N}=1$.

\subsection{Gauge symmetry breaking using the orbifold $\mathbb{T}^2/(\mathbb{Z}_2\times\mathbb{Z}_2^{\mathrm{SM}})$}
\label{orbifoldsection2}

The zero modes obtained from the compactification on
$\mathbb{T}^2/\mathbb{Z}_2$ form a $\mathcal{N}=1$ SUSY SU(5)
theory in 4d. The breaking of the $SU(5)$ gauge group
down to that of the Standard Model can be
achieved by another orbifolding. We make a coordinate shift to a new set of coordinates:
\begin{equation}
(x'_5,x'_6)=(x_5+\pi R_1,x_6)
\end{equation}
and introduce a second parity $ \mathbb{Z}^{\mathrm{SM}}_2$ on these new coordinates
\begin{equation}
\mathbb{Z}^{\mathrm{SM}}_2 : (x'_5,x'_6)\rightarrow(-x'_5,-x'_6).
\end{equation}
By using a single parity $P_{SM}$,
\begin{equation}
 P_{SM}=\begin{pmatrix}
         +1 & 0 & 0 & 0 & 0 \\
          0 & +1 & 0 & 0 & 0 \\
          0 & 0 & -1 & 0 & 0 \\
          0 & 0 & 0 & -1 & 0 \\
          0 & 0 & 0 & 0 & -1 \\
        \end{pmatrix}
\end{equation}
we shall require that:
\begin{equation}
 P_{SM}V_\mu(x,-x_5+\pi R_1/2,-x_6)P^{-1}_{SM}=+V_\mu((x,x_5+\pi R_1/2,x_6).
\end{equation}
Gauge boson fields of the standard model thus have positive parity
and fields belonging to $SU(5)/G_{SM}$ have negative parity. The
orbifold is now
$\mathbb{T}^2/(\mathbb{Z}_2\times\mathbb{Z}^{SM}_2)$.

Explicitly the expansion for the fields with any combination of
parities is:
\begin{eqnarray}
 \Phi_{++}(x,x_5,x_6)&=&\frac{1}{\pi\sqrt{R_1R_2\sin\theta}}\sum_{m\geq 0}\frac{1}{2^{\delta_{m,0}\delta_{n,0}}}\phi_{++}^{(2m,n)}(x)\nonumber\\
  & &\mbox{} \times\cos\left(\frac{2m}{R_1}\{x_5-\frac{x_6}{\tan\theta}\}+\frac{nx_6}{R_2\sin\theta} \right)\\
 \Phi_{+-}(x,x_5,x_6)&=&\frac{1}{\pi\sqrt{R_1R_2\sin\theta}}\sum_{m\geq 0}\phi_{+-}^{(2m+1,n)}(x)\nonumber\\
  & &\mbox{} \times\cos\left(\frac{(2m+1)}{R_1}\{x_5-\frac{x_6}{\tan\theta}\}+\frac{nx_6}{R_2\sin\theta} \right)\\
 \Phi_{--}(x,x_5,x_6)&=&\frac{1}{\pi\sqrt{R_1R_2\sin\theta}}\sum_{m\geq 0}\phi_{--}^{(2m,n)}(x)\nonumber\\
  & &\mbox{} \times\sin\left(\frac{2m}{R_1}\{x_5-\frac{x_6}{\tan\theta}\}+\frac{nx_6}{R_2\sin\theta} \right)\\
 \Phi_{-+}(x,x_5,x_6)&=&\frac{1}{\pi\sqrt{R_1R_2\sin\theta}}\sum_{m\geq 0}\phi_{-+}^{(2m+1,n)}(x)\nonumber\\
  & &\mbox{} \times\sin\left(\frac{(2m+1)}{R_1}\{x_5-\frac{x_6}{\tan\theta}\}+\frac{nx_6}{R_2\sin\theta} \right) .
\end{eqnarray}
Only fields with both parities positive have zero modes.

\subsection{Higgs and doublet-triplet splitting}

So far we have just considered the gauge sector of SUSY $SU(5)$.
Adding the MSSM Higgs to the 6d SUSY theory is straightforward.
In the $SU(5)$ GUT theory these are contained in
the $\textbf{5}$-plet and $\overline{\textbf{5}}$-plet of Higgs
fields. These are two complex scalars $H$ and $H'$, and a fermion
$h=(h,h')$. The chiralities are $\gamma_5h=h, \gamma_5h'=-h'$ in
4d with an overall positive 6d chrality $\Gamma_7h=h$.

The Lagrangian reads:
\begin{equation}
\mathcal{L}_{6d}^{\mathrm{higgs}}=|D_MH|^2 +
|D_MH'|^2-\frac{1}{2}g^2(H^\dagger t^a H +H'^\dagger t^a H')^2 +
i\overline{h}\Gamma^MD_Mh-i\sqrt{2}g(\overline{h}\Lambda H
+\overline{h}\Lambda^cH' + c.c).
\end{equation}
Again we integrate over the compact dimensions to get,
\begin{eqnarray}
\mathcal{L}_{4d}^{\mathrm{higgs}}& =& \sum_{m,n} i\overline{h}^{(m,n)}\gamma^\mu\partial_\mu h^{(m,n)} +i\overline{h'}^{(m,n)}\gamma^\mu\partial_\mu h'^{(m,n)}\\
& & +(\frac{m}{R_1}-i\left(\frac{n}{R_2\sin\theta}-\frac{m}{R_1\tan\theta}\right))\overline{h}^{(m,n)}h'^{(m,n)} + c.c. \\
& & +\partial_\mu H^{(m,n)\dagger} \partial^\mu H^{(m,n)} + M(m,n)^2 H^{(m,n)\dagger}H^{(m,n)} \\
& & +\partial_\mu H'^{(m,n)\dagger} \partial^\mu H'^{(m,n)} +
M(m,n)^2 H'^{(m,n)\dagger}H'^{(m,n)}.
\end{eqnarray}
For the first orbifolding parity we choose
\begin{eqnarray*}
 PH(x,-x_5,-x_6)&=&+H(x,x_5,x_6)\\
 PH'(x,-x_5,-x_6)&=&+H'(x,x_5,x_6)
\end{eqnarray*}
with $P=I$.

For the gauge breaking orbifold we choose:
\begin{eqnarray*}
 P_{SM}H(x,-x_5+\pi R_1/2,-x_6)&=&H(x,x_5+\pi R_1/2,x_6)\\
P_{SM}H'(x,-x_5+\pi R_1/2,-x_6)&=&H'(x,x_5+\pi R_1/2,x_6)
\end{eqnarray*}
It is easy to see with the form of $P_{SM}$ that the last three
entries gain a minus sign which makes them heavy whereas the first
two entries are left unchanged leaving them light, resulting in a
light doublet and a heavy coloured triplet.

\section{$A_4$ Family Symmetry from 6d $SU(5)$ SUSY GUTs}
\label{modelsection}

The model will involve an $A_4$ family symmetry which is not
assumed to exist in the 6d theory, but which originates after the
compactification down to 4d. The way this happens is quite similar to
the discussion in \cite{Altarelli:2006kg} based on the orbifold
$\mathbb{T}^2/(\mathbb{Z}_2)$ but differs somewhat due to the
different orbifold considered here, namely $\mathbb{T}^2/(\mathbb{Z}_2\times\mathbb{Z}_2^{\mathrm{SM}})$.
This is discussed in Appendix \ref{a4appendix},
where we also briefly summarize all the results required in order to formulate our model, as
necessary in order to make this paper self-contained.
Using the formalism of the previous section and Appendix \ref{a4appendix},
we now present the model.

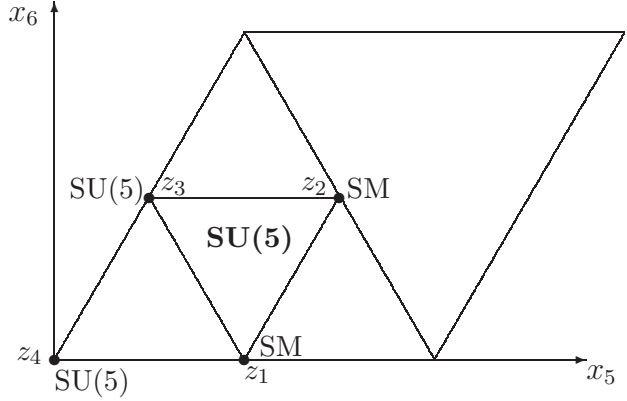
\begin{figure}
\ifx\JPicScale\undefined\def\JPicScale{.5}\fi \unitlength
\JPicScale mm
\begin{picture}(160,95)(-50,-30)
\linethickness{0.1mm} \put(10,0){\line(1,0){100}}
\put(60,87){\line(1,0){100}}
\multiput(10,0)(0.12,0.208){417}{\line(0,1){0.2}}
\multiput(110,0)(0.12,0.208){417}{\line(0,1){0.21}}
\multiput(60,87)(0.12,-0.208){417}{\line(0,-1){0.21}}
\multiput(35,43)(0.12,-0.208){209}{\line(0,1){0.21}}
\put(35,43){\line(1,0){50}}
\multiput(60,0)(0.12,0.208){209}{\line(0,1){0.21}}
\multiput(10,0)(50,0){2}{\circle*{3}}
\multiput(35,43)(50,0){2}{\circle*{3}}
\linethickness{0.1mm} \put(10,0){\vector(1,0){140}}
\put(10,0){\vector(0,1){95}}
\put(150,-5){$x_5$} \put(-2,90){$x_6$}
\small \put(10,-8){SU(5)} \put(64,1){SM} \put(14,43){SU(5)}
\put(87,43){SM}
\put(50,30){\textbf{SU(5)}} \small
\put(38,45){$z_3$}
\put(0,0){$z_4$}
\put(60,-5){$z_1$}
\put(75,45){$z_2$}
\end{picture}
\caption{\small The orbifold compactification of a 6d
$\mathcal{N}=1$ SUSY $SU(5)$ GUT which gives rise to an effective
4d theory with the $\mathcal{N}=1$ SUSY SM gauge group together
with $A_4$ Family Symmetry after compactification. The gauge
symmetry at the four fixed points is explicitly labelled. Matter fields
are localised at the fixed points as discussed in Appendix \ref{a4appendix}
and in \cite{Altarelli:2006kg}.}
\label{modelpicture}
\end{figure}

The basic set-up of the model
is depicted in Fig.~\ref{modelpicture} and the essential features
may be summarized as follows.
The model assumes a 6d gauge $\mathcal{N}=1$ SUSY $SU(5)$
Yang-Mills theory compactified down to 4d Minkowski space with two
extra dimensions compactified on a twisted torus with a twist
angle of $\theta = 60^{\circ}$ and $R_1=2R_2$. Upon compactification, without
orbifolding, the 6d supersymmetry would become extended to
$\mathcal{N}=2$ SUSY in 4d. However the $\mathcal{N}=2$ SUSY is
reduced to $\mathcal{N}=1$ SUSY by use of a particular orbifolding
and a further orbifolding is used to break the gauge symmetry to
the SM, as discussed in Section 2. Due to the tetrahedral pattern
of fixed points on the torus, the compactified extra dimensions
have some additional symmetry left over from the 6d Poincar\'{e}
spacetime symmetry, which is identified as a Family Symmetry
corresponding to the $A_4$ symmetry group of the tetrahedron. The
particular gauge breaking orbifolding also leads to the
$\mathbf{5}$-plets of higgs splitting into a light doublet and
heavy coloured triplet. It should be noted that the four fixed points
of the tetrahedral orbifold are inequivalent in that they have different gauge groups associated with them.
The $A_4$ symmetry is a symmetry of the standard model gauge bosons only and
not the full SU(5) gauge group. The gauge bosons belonging to $SU(5)/G_{SM}$ have negative parity under the
second gauge breaking orbifolding so these fields do not transform as trivial singlets under the $A_4$ as the
standard model gauge bosons do. The model is therefore $A_4\times \mathrm{SM}$ not $A_4 \times \SU(5)$.

The model is further specified by matter fields located on the
3-branes in various configurations, at the fixed points shown in
Fig.~\ref{modelpicture}. These matter fields are 4d fields with
components at the 4 fixed points as described in \cite{Altarelli:2006kg}.
Matter fields carry an extra U(1) family
dependent charge which is in turn broken by two $A_4$ singlet
Froggatt-Nielsen flavons $\theta,\theta'$ which live on the fixed
points. Realistic charged fermion masses and mixings are produced
using these Froggatt-Nielsen flavons $\theta,\theta'$ together
with the bulk flavon $\varphi_T$ which breaks $A_4$ but preserves
the $T$ generator. Tri-bimaximal mixing of the neutrinos is
achieved using further bulk flavons $ \varphi_S$ which breaks
$A_4$ but preserves the $S$ generator, and the singlet bulk flavon
$\xi$. A full list of the particle content of the model is given
in Table~\ref{particlecontent}.
The superpotential of the theory is a sum of a bulk term depending on bulk fields,
plus terms localised at the four fixed points. The 4D superpotential is produced
from the 6D theory by integrating over the extra dimensions and assuming a constant
background value for the bulk supermultiplets $\varphi_S(z),\varphi_T(z)$ and $\xi_S(z)$
as in ref \cite{Altarelli:2006kg}.

\subsection{Superfield Content}

\begin{table}[h]
 \begin{tabular}{|c|c|c|c|c|c|c|c|c|c|c|c|c|}
   \hline
   Superfield & N & F & $T_1$ & $T_2$ & $T_3$ & $H_5$ & $H_{\overline{5}}$ & $\varphi_T$&$\varphi_S$ &$\xi,\tilde{\xi}$&$\theta$&$\theta'$ \\
    \hline
    SU(5) & 1 & $\overline{5}$ & 10 & 10 & 10 & 5 & $\overline{5}$& 1 & 1 & 1 & 1 & 1 \\
    SM & 1 & $\scriptstyle(d^c,l)$& $\scriptstyle({u''}^c_1,{q''}_1,{e''}^c_1)$& $\scriptstyle({u'}^c_2,{q'}_2,{e'}^c_2)$&$\scriptstyle (u^c_3,q_3,e^c_3)$& $H_u$ &$H'_d$ &$\varphi_T$ &$\varphi_S$ & $\xi,\tilde{\xi}$& $\theta$&$\theta'$\\
    $A_4$ & 3 & 3 & $1''$ & $1'$ & 1 & 1 & $1'$ & 3 & 3 & 1 & 1 & $1'$ \\
    $U(1)$ & 0 & 0 & 4 & 2 & 0 & 0 & 0 & 0 & 0 & 0 & -1 & -1 \\
    $\mathbb{Z}_3$ & $\omega$ & $\omega$ & $\omega$& $\omega$& $\omega$& $\omega$& $\omega$ & 1 & $\omega$& $\omega$ & 1 & 1 \\
   $U(1)_R$ & 1 & 1& 1& 1& 1& 0 & 0 & 0 & 0 & 0 & 0 & 0\\
   \hline
  Brane/bulk & \tiny brane &\tiny brane &\tiny brane&\tiny brane &\tiny brane&\tiny bulk &\tiny bulk &\tiny bulk &\tiny bulk &\tiny bulk &\tiny brane &\tiny brane  \\
  \hline
   \end{tabular}
\caption{\small Superfield content and their transformation
properties under the symmetries of the model. Note that the
$SU(5)$ GUT symmetry is broken by the compactification, while the
$A_4$ Family Symmetry is only realized after the compactification.
The matter fields are located at the fixed points on 3-branes,
while the Higgs fields live in the 6d bulk. The Froggatt-Nielsen
flavons are all located at the fixed point 3-branes while the
$A_4$ flavons all live in the bulk.} \label{particlecontent}
\end{table}

After compactification, an effective 4d superpotential may be
written down, using the dictionary for the realisation of the 4d
terms in terms of the local 6d $A_4$ invariants given in
Table~\ref{dictionary}. Using this dictionary, we decompose the
effective 4d superpotential into several parts:
\begin{equation}
\label{superpotential}
 w=w_{up}+w_{down}+w_{\mathrm{charged\ lepton}}+w_{\nu}+w_{d}+\dots
\end{equation}
The term $w_d$ is concerned with vacuum alignment whose effect
will be discussed later. The first three terms give rise to the
fermion masses after $A_4$, $U(1)$ and electroweak symmetry
breaking and they are:
\begin{eqnarray}
 w_{\mathrm{up}}&\sim&\frac{1}{\Lambda}H_uq_3u^c_3+\frac{{\theta'}^2}{\Lambda^{3}}H_u(q'_2u^c_3+q_3{u'}^c_2)+\frac{{\theta'}^4+{\theta'}\theta^3}{\Lambda^{5}}H_uq_2'{u'}^c_2\nonumber\\
&+& \frac{{\theta'}^4+{\theta'}\theta^3}{\Lambda^{5}}H_u(q''_1u^c_3+q_3{u''}^c_1)+ \frac{{\theta'}^6+{\theta'}^3\theta^3+\theta^6}{\Lambda^{7}}H_u(q'_2{u''}^c_1+q''_1{u'}^c_2) \nonumber \\
&+&\frac{{\theta'}^8+{\theta'}^5\theta^3+{\theta'}^2\theta^6}{\Lambda^{9}}H_uq''_1{u''}^c_1
\end{eqnarray}

\begin{eqnarray}
 w_{\mathrm{down}}&\sim&\frac{1}{\Lambda^{3}}H'_d(d^c\varphi_T)''q_3+\frac{{\theta'}^2}{\Lambda^{5}}H'_d(d^c\varphi_T)''q'_2 + \frac{{\theta}^2}{\Lambda^{5}}H'_d(d^c\varphi_T)'q'_2 \nonumber \\
         &+& \frac{\theta'\theta}{\Lambda^{5}}H'_d(d^c\varphi_T)q'_2 +\frac{{\theta'}^4+\theta'\theta^3}{\Lambda^{7}}H'_d(d^c\varphi_T)''q''_1\nonumber \\
         &+& \frac{{\theta'}^2\theta^2}{\Lambda^{7}}H'_d(d^c\varphi_T)'q''_1+\frac{{\theta'}^3\theta+\theta^4}{\Lambda^{7}}H'_d(d^c\varphi_T)q''_1
\end{eqnarray}
\begin{eqnarray}
 w_{\mathrm{charged\ lepton}}&\sim& \frac{1}{\Lambda^{3}}H'_d(l\varphi_T)''e^c_3 +\frac{{\theta'}^2}{\Lambda^{5}}H'_d(l\varphi_T)''{e^c}'_2+\frac{{\theta}^2}{\Lambda^{5}}H'_d(l\varphi_T)'{e^c}'_2 \nonumber \\
 &+& \frac{\theta'\theta}{\Lambda^{5}}H'_d(l\varphi_T){e^c}'_2+\frac{{\theta'}^4+\theta'\theta^3}{\Lambda^{7}}H'_d(l\varphi_T)''{e^c}''_1 \nonumber \\
 &+& \frac{{\theta'}^2\theta^2}{\Lambda^{7}}H'_d(l\varphi_T)'{e^c}''_1+\frac{{\theta'}^3\theta+\theta^4}{\Lambda^{7}}H'_d(l\varphi_T){e^c}''_1
\end{eqnarray}

The dimensionless coefficients of each term in the superpotential
have been omitted and they aren't predicted by the flavour
symmetry, though they are all expected to be of the same order. It should
be noted that the up mass matrix $m_u$ is not symmetric since the lagranian is 
invariant under the standard model and not $SU(5)$.
The powers of the cut-off $\Lambda$ are determined by the
dimensionality of the various fields, recalling that brane fields
have mass dimension 1 and bulk fields have mass dimension 2 in 6d.

The neutrinos have both Dirac and Majorana masses:
\begin{equation}
 w_{\nu}\sim \frac{y^D}{\Lambda}H_u(Nl) + \frac{1}{\Lambda}(x_a\xi +\tilde{x}_a\tilde{\xi})(NN)+\frac{x_b}{\Lambda}(\varphi_S NN)
\end{equation}
where $\tilde{\xi}$ is a linear combination of two independent
$\xi$ type fields which has a vanishing VEV and therefore doesn't
contribute to the neutrino masses.

\begin{table}
\begin{center}
   \begin{tabular}{||c|c||}
   \hline
    4d & 6d \\
    \hline
    $H_uq_3u^c_3$& $\sum_i {q_3}_i{u^c_3}_i\mathbf{H}_u(z)\delta_i $ \\
   \hline
   $\theta^6\theta'^2H_uq''_1{u^c}''_1$&$\sum_i \theta^6_i\theta'^2_i\mathbf{H}_u(z){q''_1}_i{{u^c}''_1}_i\delta_i $\\
   \hline
   $\theta'^4H_uq'_2{u^c}'_2$&$\sum_i \theta'^4_i\mathbf{H}_u(z){q'_2}_i{{u^c}'_2}_i\delta_i $\\
   \hline
   $\theta'^8H_uq''_1{u^c}''_1$&$\sum_i \theta'^8_i\mathbf{H}_u(z){q''_1}_i{{u^c}''_1}_i\delta_i $\\
   \hline
   $\theta^3\theta'^3H_uq'_2{u^c}''_1$&$\sum_i \theta^3_i\theta'^3_i\mathbf{H}_u(z){q'_2}_i{{u^c}''_1}_i\delta_i $\\
   \hline
   $\theta'^4H_uq''_1{u^c}_3$&$\sum_i \theta'^4_i\mathbf{H}_u(z){q_1}''_i{{u^c}_3}_i\delta_i $\\
  \hline
  $ \theta^4H'_d({d^c}\varphi_T)q''_1$&$ \sum_{iK} \theta^4_i\mathbf{H}'_d(z)({{d^c}^{\mathcal{R}_0}}_i\alpha_{iK}{\varphi_T}_K(z)){q''_1}_i$\\
 \hline
   $ \theta^2\theta'^2 H'_d({d^c}\varphi_T)'q''_1$&
    $ \sum_{iK}\theta^2_i\theta'^2_i\mathbf{H}'_d(z)({{d^c}^{\mathcal{R}_0}}_i\alpha'_{iK}{\mathbf{\varphi}_T}_K(z))'{q''_1}_i\delta_i$\\
  \hline
  $ \theta\theta' H'_d({d^c}\varphi_T)q'_2$&
    $ \sum_{iK}\theta_i\theta'_i\mathbf{H}'_d(z)({{d^c}^{\mathcal{R}_0}}_i\alpha_{iK}{\mathbf{\varphi}_T}_K(z)){q'_2}_i\delta_i$\\
  \hline
$ H'_d({d^c}\varphi_T)''q_3$&
    $ \sum_{iK}\mathbf{H}'_d(z)({{d^c}^{\mathcal{R}_0}}_i\alpha''_{iK}{\mathbf{\varphi}_T}_K(z))''{q_3}_i\delta_i$\\
  \hline
$H_u(Nl)$ & $\sum_{i}\mathbf{H}_u(z)(N^{\mathcal{R}_0}_il^{\mathcal{R}_0}_i)\delta_i$\\
\hline
$ \xi(NN)$&$\sum_i \xi(z)(N^{\mathcal{R}_0}_iN^{\mathcal{R}_0}_i)\delta_i $\\
\hline
$ \varphi_S(NN)$&$\sum_{iK} {\varphi_S}_K(z)\alpha_{iK}N^{\mathcal{R}_0}_iN^{\mathcal{R}_0}_i\delta_i $\\
\hline
   \end{tabular}
\caption{\small A dictionary for the realisation of the 4d terms
in the superpotential in terms of the local 6d $A_4$ invariants.
The 4d terms are obtained by integrating out the extra dimensions
and assuming a constant background value for the bulk multiplets, as discussed in Appendix
\ref{a4appendix} where the notation is defined. The delta function, $\delta_i=\delta(z-z_i)$ 
where $z_i$ are the fixed points, restricts the couplings to the fixed points.}
 \label{dictionary}
\end{center}
\end{table}

Using the alignment mechanism in \cite{Altarelli:2008bg}, the
scalar components of the supermultiplets will be assumed to obtain
VEVs according to the following scheme:
\begin{eqnarray}
\frac{\braket{\varphi_{T}}}{\Lambda}&=&\frac{1}{\sqrt{\pi^2R_1R_2\sin\theta}}(v_T,0,0),\\ \frac{\braket{\varphi_S}}{\Lambda}&=&\frac{1}{\sqrt{\pi^2R_1R_2\sin\theta}}(v_S,v_S,v_S),\\
 \frac{\braket{\xi}}{\Lambda}&=&\frac{1}{\sqrt{\pi^2R_1R_2\sin\theta}}u, \\
\frac{\braket{\theta}}{\Lambda_i}&=&t_i,\\
 \frac{\braket{\theta'}}{\Lambda_i}&=&t_i'
\end{eqnarray}
where $i=u,d,e$ allowing for different messenger masses \cite{deMedeirosVarzielas:2005ax}.
Since the brane fields live in 4 dimensions the messengers will also be 4 dimensional particles
so that the mechanism in \cite{deMedeirosVarzielas:2005ax}, allowing different messenger masses, can be applied
in this scenario.
Also recall that the dimensions of the torus are now fixed
\begin{equation}
\label{fixedradii}
 R_1=2R_2 \ \ \ \mathrm{and} \ \ \ \ \sin\theta=\sqrt{3}/2.
\end{equation}
In the remainder of this paper we shall give results in terms of
$R_1$,$R_2$ and $\sin\theta$. It should be noted that they are however fixed to the values
in Eqn. (\ref{fixedradii}). Note that the flavon VEVs $v_T,v_S$ and $u$ are defined to be
dimensionless since the bulk fields have mass dimension of 2.

\subsection{Higgs vevs}
The Higgs multiplets live in the bulk this gives the required doublet-triplet splitting.
The value of the Higgs VEVs at the fixed points is what will enter
in the Yukawa couplings, so the values of we are interested in
will be averages over the fixed points $z_i$:
\begin{equation}
 \braket{\sum_iH_u(z_i)}=\frac{v_u}{\sqrt{\pi^2 R_1R_2\sin\theta}},\braket{\sum_iH'_d(z_i)}=\frac{v_d}{\sqrt{\pi^2 R_1R_2\sin\theta}}
\end{equation}
where $v_u$ and $v_d$ have mass dimension 1. The electroweak scale
will be determined by:
\begin{eqnarray}
 \mathbf{v}_u^2+\mathbf{v}_d^2 & \approx & (174GeV)^2,\\
 \mathbf{v}_u^2 & \equiv & \int d^2z|\braket{H_u(z)}|^2,\\
\mathbf{v}_d^2 & \equiv & \int
d^2z|\braket{H'_d(z)}|^2.
\end{eqnarray}
Because we are using an extra dimensional setup a suppression
factor $s$ will enter into our mass matrices since a bulk field
and it's zero mode are given by:
\begin{equation}
 \mathbf{B}=\frac{1}{\sqrt{\pi^2R_1R_2\sin\theta}}B^0+\{\mathrm{higher\ order\ contributions}\}
\end{equation}
which results in the suppression factor:
\begin{equation}
 s=\frac{1}{\sqrt{\pi^2R_1R_2\new{\sin\theta}\Lambda^2}}< 1.
\end{equation}
$R_1$,$R_2$ and $\sin\theta$ are given by Eqn. (\ref{fixedradii}). The size of s is discussed below in sec. \ref{upsector}.
%
%

\subsection{Quark and Lepton Mass Matrices}
We can now calculate the fermion mass matrices from the
effective 4d superpotential, using the flavon and Higgs VEVs and expansion parameters above,
(using a left-right convention throughout):
\begin{equation}
 m_u\sim\begin{pmatrix}
      t_u^6{t_u'}^2+{t_u'}^8+t_u^3{t_u'}^5 & t_u^6+t_u^3{t_u'}^3+{t_u'}^6 & t_u't_u^3+{t_u'}^4 \\
      t_u^6+t_u^3{t_u'}^3+{t_u'}^6 & t_u^3{t_u'}+{t_u'}^4 & {t_u'}^2\\
      t_u't_u^3+{t_u'}^4 & {t_u'}^2 &  1
     \end{pmatrix}sv_u,
\end{equation}
\begin{equation}
\label{downmatrix}
m_d\sim\begin{pmatrix}
      t_d^4+{t_d'}^3t_d & t_d^2{t_d'}^2 & t_d^3t_d'+{t_d'}^4\\
      t_dt_d' & t_d^2 & {t_d'}^2 \\
      \dots & \dots & 1
     \end{pmatrix}s^2v_Tv_d,
\end{equation}
\begin{equation}
 m_e\sim\begin{pmatrix}
      t_e^4+{t_e'}^3t_e   &   t_et_e'  & \dots  \\
      t_e^2{t_e'}^2 & t_e^2 & \dots  \\
      t_e^3t_e'+{t_e'}^4 & {t_e'}^2 & 1
     \end{pmatrix}s^2v_Tv_d,
\end{equation}
where we have achieved different values for $t_u,t_d$ and $t_e$ via different messenger masses $\Lambda_u,\Lambda_d$ and $\Lambda_e$
and the dots represent contributions from subleading operators as discussed in sec. \ref{alignmentsection}.
\subsubsection{Down sector}
For the down quark mass matrix, $m_d$, we can choose $t_d\sim \epsilon$ and $t_d'\sim\epsilon^{2/3}$ to give:
\begin{equation*}
 m_d\sim\begin{pmatrix}
      \epsilon^3 & \epsilon^{10/3} & \epsilon^{8/3} \\
      \epsilon^{5/3} & \epsilon^2 & \epsilon^{4/3} \\
      \dots & \dots & 1
     \end{pmatrix}v_T s^2 v_d.
\end{equation*}
For example, assuming a value $\epsilon \approx 0.15$ allows the
order unity coefficients to be tuned to $\mathcal{O}(\epsilon)$ to give acceptable
down-type quark mass ratios. The 11 element of the mass matrix is of order $\epsilon^3$, which
needs to be tuned to order $\epsilon^4$ using the dimensionless coefficients we have omitted to write
in the superpotential. The dots again represent subleading operators as discussed in
sec. \ref{alignmentsection}.

\subsubsection{Up sector}
\label{upsector}
The up quark matrix is given by:
\begin{equation*}
 m_u\sim\begin{pmatrix}
                  \bar{\epsilon}^8 & \bar{\epsilon}^{6} & \bar{\epsilon}^4 \\
                  \bar{\epsilon}^{6} & \bar{\epsilon}^4 & \bar{\epsilon}^2 \\
                  \bar{\epsilon}^4 & \bar{\epsilon}^2 & 1 \\
                   \end{pmatrix}sv_u
\end{equation*}
with $t_u\sim t_u'\sim \bar{\epsilon}$ . Again we have left out
the $\mathcal{O}(1)$ coefficients for each term, which for
$\bar{\epsilon} \approx 0.22$, may be tuned to give acceptable
up-type quark mass ratios. The CKM mixing angles will arise
predominantly from the down-mixing angles, but with possibly
significant corrections from the up-mixing angles, depending on
the unspecified operators represented by dots. In general there
will be corrections to all the Yukawa matrices as discussed later.
Since the top mass is given by the size of $s$, we would expect a value
around $s\sim0.5$.
\subsubsection{Charged lepton mass matrix}
The mass matrix for the charged lepton sector is of the form:
\begin{equation*}
 m_e\sim\begin{pmatrix}
      t_e^4 + {t_e'}^3t_e   &   t_et_e'  & \dots  \\
      t_e^2{t_e'}^2 & t_e^2 & \dots  \\
      t_e^3t_e'+{t_e'}^4 & {t_e'}^2 & 1
      \end{pmatrix}s^2v_Tv_d=\begin{pmatrix}
                               \epsilon^3 & \epsilon^{5/3} & \dots \\
                               \epsilon^{10/3} & \epsilon^2 & \dots \\
                               \epsilon^{8/3} & \epsilon^{4/3} & 1
                               \end{pmatrix}v_T s^2 v_d.
\end{equation*}
with $t_e\sim\epsilon$ and $t_e'\sim \epsilon^{2/3}$. The dots again represent subleading
operators as discussed in sec. \ref{alignmentsection}.
%

\subsubsection{Neutrino sector}
The neutrino sector after the fields develop VEVs and the gauge
singlets N become heavy the see-saw mechanism takes place as
discussed in detail in \cite{Chen:2009um}. After the see-saw mechanism the effective mass matrix for the
light neutrinos is given by:
\begin{equation}
 m_\nu\sim\frac{1}{3a(a+b)}\begin{pmatrix}
                         3a+b & b & b \\
                        b & \frac{2ab+b^2}{b-a} & \frac{b^2-ab-3a^2}{b-a}\\
                        b & \frac{b^2-ab-3a^2}{b-a} & \frac{2ab+b^2}{b-a}
                        \end{pmatrix}\frac{{s(v_u)^2}}{\Lambda}
\end{equation}
where
\begin{equation*}
 a\equiv\frac{2x_au}{(y^D)^2},b\equiv \frac{2x_bv_S}{(y^D)^2}
\end{equation*}
The neutrino mass matrix is diagonalised by the transformation
\begin{equation*}
 U_{\nu}^Tm_\nu U_{\nu}=\mathrm{diag}(m_1,m_2,m_3)
\end{equation*}
with $U_{\nu}$ given by:
\begin{equation*}
 U_{\nu}=\begin{pmatrix}
    -\sqrt{2/3} & 1/\sqrt{3} & 0 \\
     1/\sqrt{6} & 1/\sqrt{3} & 1/\sqrt{2} \\
     1/\sqrt{6} & 1/\sqrt{3} & -1/\sqrt{2}
   \end{pmatrix}
\end{equation*}
which is of the TB form in Eq.~(\ref{MNS0}). However, although we
have TB neutrino mixing in this model we do not have exact TB
lepton mixing due to fact that the charged lepton mass matrix is
not diagonal in this basis. Thus there will be charged lepton
mixing corrections to TB mixing resulting in mixing sum rules as
discussed in \cite{King:2005bj,Antusch:2008yc}.

\section{Vacuum alignment and subleading corrections}
\label{alignmentsection}

\begin{table}[h]
\begin{center}
 \begin{tabular}{|c|c|c|c|c|c|c|c|}
  \hline
  Field & $\varphi_T$ & $\varphi_S$ & $\xi$ & $\tilde{\xi}$ & $\varphi^T_0$ & $\varphi^S_0$ &$\xi_0$ \\
\hline
  $\mathbb{Z}_3$ & 1&$\omega$ & $\omega$& $\omega$&1 &$\omega$ &$\omega$ \\
\hline
 $U(1)_R$ & 0 & 0& 0 & 0& 2 & 2 & 2\\
\hline
 Brane/Bulk & \small{Bulk}&\small{Bulk} &\small{Bulk}&\small{Bulk}&\small{Bulk}&\small{Bulk}&\small{Bulk}\\
\hline
 \end{tabular}
\label{drivingfields}
\caption{\small The flavon fields and driving fields leading to the vacuum alignment.}
\end{center}
\end{table}

The resulting $A_4$ model is of the direct kind discussed in
\cite{King:2009ap} in which the vacuum alignment is achieved via
F-terms resulting in the $A_4$ generator $S$ being preserved in
the neutrino sector. The vacuum alignment is achieved by the
superpotential $w_d$ introduced in \cite{Altarelli:2008bg}, where
we have absorbed the mass dimension into the coefficients  $g_i,f_i$.
\begin{eqnarray}
 w_d&=& M(\varphi_T\varphi_0^T)+g(\varphi_0^T\varphi_T\varphi_T)+g_1(\varphi_0^S\varphi_S\varphi_S) \nonumber\\
   & +& (f_1\xi+f_2\tilde{\xi})\varphi_0^S\varphi_S + f_3\xi_0(\varphi_S\varphi_S) \nonumber \\
   &+& f_4\xi_0\xi\tilde{\xi}+f_5\xi_0\xi^2+f_6\xi_0\tilde{\xi}^2,
\end{eqnarray}
involving additional gauge singlets, the driving fields
$\varphi^T_0,\varphi^S_0$ and $\xi_0$ in
Table~3. The above form of the driving superpotential $w_d$ and
the vanishing of the F-terms,
\begin{equation}
 \frac{\partial w}{\partial \varphi^T_0}=\frac{\partial w}{\partial \varphi^S_0}=\frac{\partial w}{\partial \xi_0}=0,
\end{equation}
yields the vacuum alignment anticipated in the previous section. For more details see \cite{Altarelli:2008bg}.
Note that the FN flavons $\theta, \theta'$ require no special vacuum alignment and their VEVs may be
generated dynamically by a radiative symmetry breaking mechanism. The ratio of VEVs of
$\theta, \theta'$ will depend on the details of all the Yukawa couplings involving these flavons
from which the desired VEVs can emerge.
In general we do not address the question of the correlation of flavon VEVs in this paper.

\subsection{Subleading corrections}

Subleading corrections in the mass matrices arise from shifts in
the VEVs of the flavons, and the shifted VEVs including such
corrections are of the general form:
\begin{eqnarray}
 \braket{\varphi_T}/\Lambda &=& \frac{1}{\sqrt{\pi^2R_1R_2\sin\theta}}(v_T+\delta {v_T},\delta {v_T},\delta {v_T}) \\
 \braket{\varphi_S}/\Lambda &=& \frac{1}{\sqrt{\pi^2R_1R_2\sin\theta}}(v_S+\delta {v_S}_1,v_S+\delta {v_S}_2,v_S+\delta {v_S}_3) \\
\braket{\xi}/\Lambda&=&\frac{1}{\sqrt{\pi^2R_1R_2\sin\theta}}u\\
 \braket{\tilde{\xi}}/\Lambda&=&\frac{1}{\sqrt{\pi^2R_1R_2\sin\theta}}\delta u'
\end{eqnarray}
as discussed in \cite{Altarelli:2008bg},\cite{Altarelli:2005yx}. $\varphi_T$ obtains a correction proportional to the VEV of $\varphi_S$, where
$\varphi_S$ obtains a correction in an arbitrary direction. 
The VEV of $\tilde{\xi}$, which was zero at leading order, obtains a small
correction. The shift in the VEV of $\xi$ has been absorbed into a redefinition of $u$
since at this stage $u$ is a free parameter.

\subsection{Corrections to $m_{\mathrm{up}}$}

The leading order terms in the up sector are of the form
$\theta^m\theta'^nH_uq_iu_j$. Terms are gauge and $A_4$ singlets,
to create higher order terms we need to introduce flavon fields.
The most straighforward way to do this is to introduce two flavon
fields $\varphi_T\varphi_T$, since $\varphi_T$ is an $A_4$ triplet
we need two fields in order to construct a singlet. Such terms
will lead to entries in the mass matrix suppressed by a factor of
$v_T^2$. Because of the $\mathbb{Z}_3$ symmetry the flavon fields
$\varphi_S,\xi,\tilde{\xi}$ must enter at the three flavon level
so entries will be supressed by a factor of $v_S^2u,v_S^3$ and
$u^3$ relative to the leading order term.

\subsection{Corrections to $m_{\mathrm{down}}$ and $m_{\mathrm{charged\ lepton}}$}

In the down mass matrix subleading corrections fill in the entries
indicated by dots. Entries in the matrix are generated by terms of
the form $ \theta^m\theta'^nH'_d((d^c\varphi_T)q_i+(l\varphi_T)e^c_i)$,
higher order terms can come from replacing $\varphi_T$ with a
product of flavon fields or including the effect of the
corrections to the VEV of $\varphi_T$. We can replace $\varphi_T$
with $\varphi_T\varphi_T$, this is compatible with the
$\mathbb{Z}_3$ charges and results in corrections with the same
form as $m_{\mathrm{down}}$ but with an extra overall supression
of $v_T$. If we include the corrections to the VEV of $\varphi_T$
then we fill in the entries indicated by dots in eqn.
(\ref{downmatrix}), the corrections are of the form:


\begin{equation}
 m_d\sim\begin{pmatrix}
      \epsilon^{{8/3}}\delta v_T & \epsilon^{{8/3}}\delta v_T & \epsilon^{{8/3}}\delta v_T \\
      \epsilon^{{4/3}}\delta v_T & \epsilon^{{4/3}}\delta v_T & \epsilon^{{4/3}}\delta v_T \\
      \delta v_T & \delta v_T & \delta v_T
     \end{pmatrix} s^2 v_d .
\end{equation}
The corrections to the charged lepton mass matrix are, up to $\mathcal{O}(1)$ cooefficients, the transpose of the above matrix:
\begin{equation}
 m_e\sim\begin{pmatrix}
      \epsilon^{{8/3}}\delta v_T & \epsilon^{{4/3}}\delta v_T & \delta v_T \\
      \epsilon^{{8/3}}\delta v_T & \epsilon^{{4/3}}\delta v_T & \delta v_T \\
      \epsilon^{{8/3}}\delta v_T & \epsilon^{{4/3}}\delta v_T & \delta v_T
     \end{pmatrix} s^2 v_d .
\end{equation}

Following ref. \cite{Altarelli:2008bg}, $\delta v/v \sim \mathcal{O}(\epsilon^2)$ leading to negligible corrections to the leading order
$m_d,m_e$ mass matrices.
\subsection{Corrections to $m_{\nu}$}

The Dirac mass term ($H_u(Nl)$) can be modified with an insertion of the $\varphi_T$ flavon, producing corrections suppressed by $sv_T$.
The leading Dirac mass correction is the term $H_u(\varphi_TNl)$. This leads to a correction to the Dirac mass matrix suppressed by a factor
of $sv_T$ relative to the leading order (LO) term.

\begin{equation}
 m_{LR}=m^{\scriptscriptstyle LO}_{LR}+\Delta m_{LR}=y^Dsv_u\begin{pmatrix}
                                 1&0&0\\
                                 0&0&1\\
                                 0&1&0
                                \end{pmatrix}
                                +v_us^2v_T\begin{pmatrix}
                                  2/3& 0 & 0\\
                                   0 & 0 & 1/6\\
                                   0 & -5/6 & 0
                                  \end{pmatrix}
\end{equation}
The Majorana mass term can receive corrections from a number of higher order terms since the $(NN)$ term can be a $1,1',1''$ or 3.
The higher order terms all consist of insertions of 2 flavon fields where the leading order terms have only one insertion
e.g. the term $(NN)'(\varphi_T\varphi_S)''$ obeys the $\mathbb{Z}_3$ symmetry, is an $A_4$ singlet and results in a higher order
correction to the terms $(x_a\xi +\tilde{x}_a\tilde{\xi})(NN)+x_b(\varphi_S NN)$. If we call the correction to the Majorana
mass matrix $\delta m_{RR}$ then for this example the correction is given below,
\begin{eqnarray}
 m_{RR}&=&m^{\scriptscriptstyle LO}_{RR}+\delta m_{RR}\\
m^{\scriptscriptstyle LO}_{RR}&=&x_asu\Lambda\begin{pmatrix}
                                              1 & 0 & 0\\
                                              0 & 0 & 1 \\
                                              0 & 1 & 0
                                             \end{pmatrix}+
                                                           \frac{x_bsv_S\Lambda}{3}\begin{pmatrix}
                                                                                    2& -1 & -1 \\
                                                                                   -1 & 2 & -1 \\
                                                                                   -1 & -1 & 2
                                                                                   \end{pmatrix}\\
\delta m_{RR}&=&s^2\Lambda v_Tv_S\begin{pmatrix}
                                  0&1&0\\
                                  1&0&0\\
                                  0&0&1
                                 \end{pmatrix}.
\end{eqnarray}
Such corrections have a relative supression of $sv_{T,S}$ to the leading order term.
After the see-saw mechanism this leads to an effective mass matrix with every entry suppressed
by a factor of $sv_{T,S}$. This leads to corrections to the neutrino tri-bimaximal mixing angles of order $sv_{T,S}$.
\begin{eqnarray}
 m_{\nu}+\Delta m_{\nu}&=&m_{LR}m_{RR}^{-1}m_{LR}^t = (m_{LR}^{LO}+\Delta m_{LR})(m_{RR}^{LO}+\Delta m_{RR})^{-1}(m_{LR}^{LO}+\Delta m_{LR})^t\nonumber\\
 \frac{(\Delta m_\nu)_{ij}}{(m_\nu)_{ij}}&\sim& \mathcal{O}(sv_{T,S})
\end{eqnarray}
The magnitude of $v_T$ depends on the ratio of the top and bottom quark Yukawa couplings, but may be roughly between
$v_T\sim\mathcal{O}(\epsilon^2)-\mathcal{O}(\epsilon)$ leading to significant corrections
to tri-bimaximal mixing. The flavon shifts $\delta v_S$ also give corrections to the leading order term $(x_b(\varphi_S NN))$, however if $v_T\sim \mathcal{O}(\epsilon^2)$ 
these corrections are of $\mathcal{O}(\epsilon^2)$ and they enter at the same order of magnitude as the corrections from higher order corrections. If however 
$v_T\sim\mathcal{O}(\epsilon)$ then the correction enters at the order of $\epsilon$.The effect of the VEV of $\tilde{\xi}$, 
which was zero at leading order, and obtains a small
correction, leads to a small shift in the overall scale of the right-handed neutrino masses.
And, as already remarked, the shift in the VEV of $\xi$ has been absorbed into a redefinition of $u$, which we are free to do since $u$ is a free parameter.

\section{Conclusion}
\label{conclusion}
We have proposed a model in which an $A_4$ Family Symmetry arises
dynamically from an $\mathcal{N}=1$ $SU(5)$ SUSY GUT in 6d. The
$A_4$ Family Symmetry emerges as a result of the compactification
of the extra complex compact dimension $z$, assuming a particular
twist angle $\theta = 60^\circ$ and a particular orbifold
$\mathbb{T}^2/(\mathbb{Z}_2\times\mathbb{Z}_2^{\mathrm{SM}})$
which breaks the $\mathcal{N}=1$ $SU(5)$ SUSY GUT in 6d down to
the effective 4d $\mathcal{N}=1$ SUSY SM gauge group. In this
model the $A_4$ Family Symmetry emerges after compactification as
a residual symmetry of the full 6d spacetime symmetry of 6d
translations and proper Lorentz transformations. It should be
noted that had improper Lorentz transformations been included then
the residual symmetry would have been $S_4$ and not $A_4$. The
model also involves other symmetries, in particular we assume a
Froggatt-Nielsen $U(1)$ Family Symmetry and other $\mathbb{Z}_N$
symmetries in order to achieve a realistic model.

We emphasize that the $SU(5)$ GUT symmetry is broken by the
compactification, while the $A_4$ Family Symmetry is only realized
after the compactification. The matter fields are located at the
fixed points on 3-branes, while the Higgs fields live in the 6d
bulk. The Froggatt-Nielsen flavons are all located at the fixed
point 3-branes while the $A_4$ flavons all live in the bulk. We
have adopted an $A_4$ classification scheme of quarks and leptons
compatible with the $SU(5)$ symmetry. We have also used a
Froggatt-Nielsen mechanism for the inter-family mass hierarchies.
By placing the $\mathbf{5}$
and $\overline{\mathbf{5}}$ of Higgs in the 6d bulk we have
avoided the doublet-triplet splitting problem by making the
coloured triplets heavy. The model naturally has TB mixing at the
first approximation and reproduces the correct mass hierarchies
for quarks and charged leptons and the CKM mixing pattern. The
presence of $SU(5)$ GUTs means that the charged lepton mixing
angles are non-zero resulting in predictions such as a lepton
mixing sum rule of the kind discussed in
\cite{King:2005bj,Antusch:2008yc}.

In conclusion, this paper represents the first realistic 6d
orbifold $SU(5)$ SUSY GUT model in the literature which leads to
an $A_4$ Family Symmetry after compactification. We emphasize that
the motivation for building such higher dimensional models is
purely bottom-up, namely to make contact with high energy theories
and to solve the conceptual problems with GUT theories such as
Higgs doublet-triplet splitting and the origin of Family Symmetry
in a higher dimensional setting. The hope is that 6d models such
as the one presented here, based on one extra complex dimension
$z$, may provide a useful stepping-stone towards a 10d fully
unified string theory (including gravity, albeit perhaps decoupled
in some limit) in which GUT breaking and the emergence of Family
Symmetry can both be naturally explained as the result of the
compactification of three extra complex dimensions.

\section*{Acknowledgements}
We acknowledge partial support from the following grant: STFC
Rolling Grant ST/G000557/1.

\appendix

\section{The group $A_4$ and it's representations}

The $A_4$ group is the group of even permutations of 4 objects.
There are 4!/2=12 elements. This group can be seen as the symmetry
group of the tetrahedron, the odd permutations can be seen as the
exchange of two vertices which can't be obtained with a rigid
solid). Let a generic permutation be denoted by
$(1,2,3,4)\rightarrow(n_1,n_2,n_3,n_4)=(n_1n_2n_3n_4)$. $A_4$ can
be generated by the two basic permutations S and T where
$S=(4321)$ and $T=(2314)$. We can check that
\begin{equation*}
S^2=T^3=(ST)^3=1.
\end{equation*}
This is called the presentation of the group.

\subsection{Equivalence classes}

There are 4 equivalence classes ( $h$ and $k$ belong to the same
equivalence class if there is a member of the group $g$ such that
$ghg^{-1}=k$):
\begin{eqnarray*}
C1 :& I=(1234)\\
C2 :& T=(2314),ST(4132),TS=(3241),STS=(1423)\\
C3 :& T^2=(3124),ST^2=(4213),T^2S=(2431),TST=(1342)\\
C4 :& S=(4321),T^2ST=(3412),TST^2=(2143).
\end{eqnarray*}
For a finite group the squared dimensions for each inequivalent
representation sum to N, the number of transformations in the
group (N=12 for $A_4$). There are 4 inequivalent representations
of $A_4$ three singlets 1,$1'$,$1''$ and a triplet 3. The three
singlets representations are:
\begin{eqnarray*}
&1\hspace{3pt}& S=1 \hspace{5pt} T=1 \\
&1'\hspace{3pt}& S=1 \hspace{5pt} T=e^{2\pi i/3}=\omega \\
&1''\hspace{3pt&} S=1 \hspace{5pt} T=e^{4\pi i/3}=\omega^2.
\end{eqnarray*}
The triplet representation in the basis where S is diagonal is
constructed from:
\begin{equation*}
S=\begin{pmatrix}
    1& 0 & 0 \\
    0 & -1 & 0 \\
    0 & 0 & -1
   \end{pmatrix},\ \
T=\begin{pmatrix}
  0 & 1 & 0 \\
  0 & 0 & 1 \\
  1 & 0 & 0
  \end{pmatrix}
\end{equation*}

\subsection{Characters}
\label{characters}
The characters of a group $\chi^R_g$ of each element $g$ are
defined as the trace of the matrix that maps the element in a
representation $R$. Equivalent representations have the same
characters and the characters have the same value for all the
elements in an equivalence class. Characters satisfy $\sum_g
\chi^R_g {\chi^S_g}^*=N\delta^{RS}$. Also the character for an
element $h$ in a direct product of representations is a product of
characters $\chi^{R\otimes S}_h=\chi^R_h\chi^S_h$and is also equal
to the sum of characters in each representation that appears in
the decomposition of $R\otimes S$.

\begin{table}
\begin{center}
  {\begin{tabular}{|c|c|c|c|c|}
 \hline
   Class & $\chi_1$ & $\chi_{1'}$ & $\chi_{1''}$ & $\chi_3$ \\
   \hline
   $C_1$ & 1 & 1 & 1 & 3 \\
   \hline
   $C_2$ & 1 & $\omega$ & $\omega^2$ & 0\\
   \hline
   $C_3$ & 1 & $\omega^2$ & $\omega$ & 0 \\
   \hline
   $C_4$ & 1 & 1 & 1 & -1 \\
   \hline
  \end{tabular}}
\caption{\small The character table of $A_4$.}
\label{charactertable}
\end{center}
\end{table}
From the character table~\ref{charactertable} we can see that
there are no more inequivalent irreducible representations than
$1,1',1''$ and $3$. We can also see the multiplication rules:
\begin{eqnarray*}
3 \times 3 &=& 1 +1'+1'' + 3 + 3 \\
1'\times1'&=& 1'' \\
1'\times1''&=& 1 \\
1''\times1''&=&1'.
\end{eqnarray*}
If we have two triplets $3_a\sim (a_1,a_2,a_3)$ and
$3_b\sim(b_1,b_2,b_3)$ we can obtain the irreducible
representations from their product:
\begin{eqnarray*}
 1 &=&a_1b_1+a_2b_2+a_3b_3\\
 1'&=& a_1b_1+\omega^2a_2b_2+\omega a_3b_3\\
 1'' &=& a_1b_1+\omega a_2b_2+\omega^2a_3b_3\\
 3_{\mathrm{s}} &\sim & (a_2b_3,a_3b_1,a_1b_2)\\
 3_{\mathrm{a}} &\sim &(a_3b_2,a_1b_3,a_2b_1).
\end{eqnarray*}

\subsection{Another representation}

Previously we used the representation where the matrix S is
diagonal. In this paper we shall construct the model in a
different basis in which we arrange T to be diagonal through a
unitary transformation:
\begin{eqnarray*}
T'=VTV^\dagger=\begin{pmatrix}
               1 & 0 & 0 \\
               0 & \omega & 0 \\
               0 & 0 & \omega^2
               \end{pmatrix}, \ \
S'=VSV^\dagger=\frac{1}{3}\begin{pmatrix}
                           -1 & 2 & 2 \\
                           2 & -1 & 2 \\
                           2 & 2 & -1
                          \end{pmatrix}
\end{eqnarray*}
where
\begin{equation*}
    V=\frac{1}{\sqrt{3}}\begin{pmatrix}
                        1 & 1 & 1 \\
                        1 & \omega^2 & \omega \\
                        1 & \omega & \omega^2
                        \end{pmatrix}.
\end{equation*}
In this basis the product composition rules are different:
\begin{eqnarray}
\label{tdiagonalbasis1}
1 &=& a_1b_1+a_2b_3+a_3b_2\\
\label{tdiagonalbasis2}
1' &=& a_3b_3+a_1b_2+ a_2b_1\\
\label{tdiagonalbasis3}
1'' &=& a_2b_2+ a_1b_3+a_3b_1\\
3_{\mathrm{s}} &\sim & \frac{1}{3}(2a_1b_1-a_2b_3-a_3b_2,2a_3b_3-a_1b_2-a_2b_1,2a_2b_2-a_1b_3-a_3b_1)\\
3_{\mathrm{a}} &\sim
&\frac{1}{2}(a_2b_3-a_3b_2,a_1b_2-a_2b_1,a_1b_3-a_3b_1).
\end{eqnarray}
%
%
As discussed in Appendix \ref{a4appendix}, this is done by applying a
matrix $v=Uu$ which block diagonalises the generators of $A_4$.
This formula allows us to write triplets and singlets of our 6d
theory in terms of brane fields at the four fixed points.

\section{$A_4$ Family Symmetry from 6d compactification}
\label{a4appendix}
\new{In this Appendix we adapt the calculation in \cite{Altarelli:2006kg} to the orbifold $\mathbb{T}^2/(\mathbb{Z}_2\times\mathbb{Z}_2^{\mathrm{SM}})$.}
\subsection{The $A_4$ orbifold $\mathbb{T}^2/(\mathbb{Z}_2\times\mathbb{Z}_2^{\mathrm{SM}})$}
The Orbifold we are using is based on the twisted torus with the twist angle
$\theta=60^\circ$. We set $R_1=2R_2$, as shown in figure \ref{a4torus}, then under the orbifolding $\mathbb{Z}_2^{SM}$
the fundamental domain is reduced to a rhombus. We then perfom another orbifolding $\mathbb{Z}_2$
which folds the rhombus into a tetrahedron giving rise to the $A_4$ symmetry, as described in Appendix \ref{a4appendix}, we later exploit as a
family symmetry.
The fixed points are inequivalent but the $A_4$ symmetry is a symmetry of the Standard Model Gauge bosons only i.e.
\begin{eqnarray*}
\mathcal{S}: V^{\mathrm{SM}}_\mu(z_S)&=&V^{\mathrm{SM}}_\mu(z)\nonumber \\
\mathcal{T}: V^{\mathrm{SM}}_\mu(z_T)&=&V^{\mathrm{SM}}_\mu(z)\nonumber \\
\mathcal{S}: V^{\mathrm{SU(5)}/\mathrm{SM}}_\mu(z_S)&\neq&V^{\mathrm{SU(5)}/\mathrm{SM}}_\mu(z)\nonumber \\
\mathcal{T}: V^{\mathrm{SU(5)}/\mathrm{SM}}_\mu(z_T)&\neq&V^{\mathrm{SU(5)}/\mathrm{SM}}_\mu(z)
\end{eqnarray*}
where $z_S,z_T$ are the coordinate transformations that generate the S and T generators of the $A_4$ group. This makes explicit that the $A_4$
symmetry is a symmetry of the standard model but not SU(5).

\subsection{The orbifold with $\theta=\pi/3$}
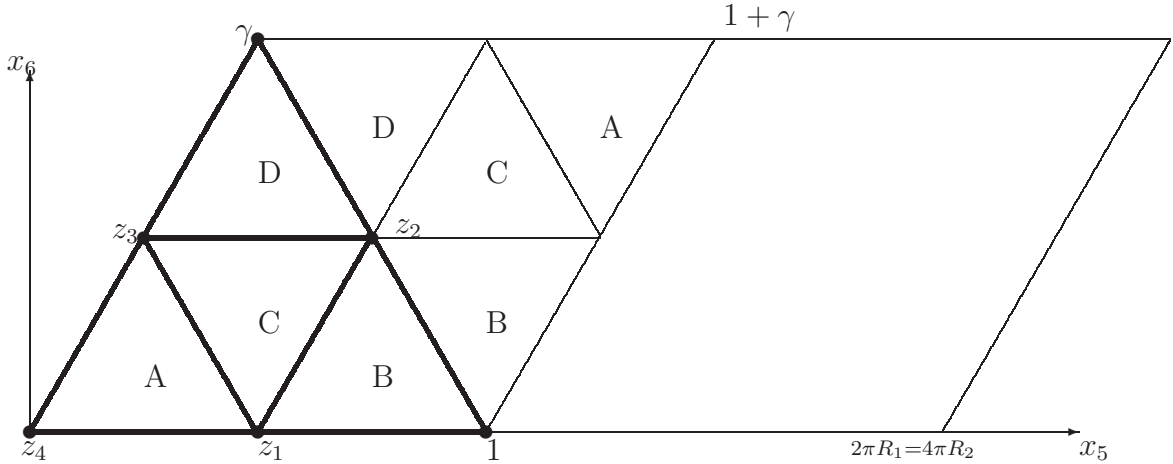
\begin{figure}
\ifx\JPicScale\undefined\def\JPicScale{.6}\fi \unitlength
\JPicScale mm
\begin{picture}(160,95)(0,-10)
\linethickness{0.6mm} \put(10,0){\line(1,0){100}}
\linethickness{0.1mm} \put(110,0){\line(1,0){100}}
\linethickness{0.1mm} \put(60,87){\line(1,0){200}}
\linethickness{0.6mm}
\multiput(10,0)(0.12,0.208){417}{\line(0,1){0.2}}
\linethickness{0.1mm}
\multiput(110,0)(0.12,0.208){417}{\line(0,1){0.21}}
\multiput(210,0)(0.12,0.208){417}{\line(0,1){0.21}}
\linethickness{0.6mm}
\multiput(60,87)(0.12,-0.208){417}{\line(0,-1){0.21}}
\linethickness{0.6mm}
\multiput(60,0)(0.12,0.208){208}{\line(0,1){0.21}}
\linethickness{0.1mm}
\multiput(85,43)(0.12,0.208){208}{\line(0,1){0.21}}
\linethickness{0.1mm}
\multiput(110,87)(0.12,-0.208){208}{\line(0,-1){0.22}}
\linethickness{0.6mm} \put(35,43){\line(1,0){50}}
\linethickness{0.1mm} \put(85,43){\line(1,0){50}}
\linethickness{0.6mm}
\multiput(35,43)(0.12,-0.208){208}{\line(0,-1){0.19}}
\multiput(10,0)(50,0){3}{\circle*{3}}
\multiput(35,43)(50,0){2}{\circle*{3}} \put(60,87){\circle*{3}}
\linethickness{0.1mm} \put(10,0){\vector(1,0){230}}
\put(10,0){\vector(0,1){80}}
\put(190,-5){$\scriptstyle 2\pi R_1=4\pi R_2$}
\put(240,-5){$x_5$} \put(5,80){$x_6$}
\put(8,-5){$z_4$} \put(60,-5){$z_1$} \put(28,43){$z_3$}
\put(90,44){$z_2$}
\put(55,87){$\gamma$} \put(162,90){$1+\gamma$} \put(110,-6){$1$}
\put(135,65){A} \put(35,10){A} \put(85,10){B} \put(110,22){B}
\put(60,22){C} \put(110,55){C} \put(60,55){D} \put(85,65){D}
\end{picture}
\caption{\small \new{The Orbifold $\mathbb{T}^2/(\mathbb{Z}_2\times\mathbb{Z}^{\mathrm{SM}}_2)$. The
fundamental domain is outlined in bold and forms a tetrahedron.
Regions labelled by A,B,C and D are identified. The fixed points
are labelled $z_i$ and are symmetrically permuted under the
symmetry group $A_4$.}}
\label{a4torus}
\end{figure}


We are working with a quantum field theory in 6 dimensions with
the 2 extra dimensions compactified onto an orbifold
$\mathbb{T}^2/(\mathbb{Z}_2\times\mathbb{Z}^{\mathrm{SM}}_2)$. The
extra dimensions are complexified such that $z=x_5+ix_6$ are the
coordinates on the extra space. The torus $\mathbb{T}^2$ is
defined by identifying the points (as in
Eq.~(\ref{twistedsymmetry}))
\new{\begin{eqnarray}
\label{zplus1}
 z\rightarrow z+2 \\
\label{zplusgamma} z\rightarrow z+\gamma  & &
\gamma=e^{i\frac{\pi}{3}}.
\end{eqnarray}}
We have set the length \new{$2\pi R_2$} to unity for clarity.
\new{If we first perform the gauge breaking orbifolding $\mathbb{Z}_2^{\mathrm{SM}}$ by making a coordinate
shift as described in sect. \ref{orbifoldsection2}
\begin{equation}
(x'_5,x'_6)=(x_5+\pi R_1,x_6)=(x_5+1,x_6)
\end{equation}
and introduce a parity $ \mathbb{Z}^{\mathrm{SM}}_2$ on these new coordinates
\begin{equation}
\mathbb{Z}^{\mathrm{SM}}_2 : (x'_5,x'_6)\rightarrow(-x'_5,-x'_6).
\end{equation}
we are left with a fundamental domain in the shape of rhombus.}
The second orbifolding is defined,as in sect. \ref{orbifoldsection}, by the parity $\mathbb{Z}_2$ identifying:
\begin{eqnarray}
\label{orbifoldparity}
 z&\rightarrow & -z\\
 (x_5,x_6)&\rightarrow& (-x_5,-x_6) \nonumber
\end{eqnarray}
leaving the orbifold to be represented by the triangular region
shown in Fig.~\ref{a4torus}. The orbifold has 4 fixed points which
are unchanged under the symmetries of the orbifold Eqns.
(\ref{orbifoldparity}),(\ref{zplus1}),(\ref{zplusgamma}). The
orbifold can be described as a regular tetrahedron with the fixed
points as the vertices. The 6d spacetime symmetry is broken by the
orbifolding, previously the symmetry consisted of 6d translations
and proper Lorentz tranformations\footnote{if we had allowed
improper lorentz transformations,i.e. reflections, then rather
than $A_4$ we would have $S_4$ the group of permutations of 4
objects}. We are now left with a 4d space-time symmetry and a
discrete symmetry of rotations and translations due to the special
geometry of the orbifold. We can generate this group with the
transformations:
\begin{eqnarray}
\mathcal{S}: z\rightarrow& z+\frac{1}{2}&\\
\mathcal{T}: z\rightarrow& \omega z &,\hspace{75pt} \omega \equiv \gamma^2
\end{eqnarray}
These two generators are even permutations of the four fixed
points:
\begin{eqnarray}
 \mathcal{S}:(z_1,z_2,z_3,z_4)\rightarrow (z_4,z_3,z_2,z_1) \\
\mathcal{T}:(z_1,z_2,z_3,z_4)\rightarrow (z_2,z_3,z_1,z_4).
\end{eqnarray}
The above two transformations generate the group $A_4$ which is
the symmetry of the tetrahedron (see Appendix A for an
introduction to $A_4$). This can be verified by showing that S and
T obey the characteristic relations, the presentation, of the
generators of $A_4$,
\begin{equation}
\label{presentation}
 \mathcal{S}^2=\mathcal{T}^3=(\mathcal{S}\mathcal{T})^3=1.
\end{equation}
We can easily represent S and T by $4\times4$ matrices describing
their action on the fixed points of the orbifold:
\begin{eqnarray}
 S=\begin{pmatrix}
    0 & 0 & 0 & 1 \\
    0 & 0 & 1 & 0 \\
    0 & 1 & 0 & 0 \\
    1 & 0 & 0 & 0
   \end{pmatrix}
   &,&
  T=\begin{pmatrix}
    0 & 1 & 0 & 0 \\
    0 & 0 & 1 & 0 \\
    1 & 0 & 0 & 0 \\
    0 & 0 & 0 & 1
   \end{pmatrix}.
\end{eqnarray}

The 4d representations of the $A_4$ generators \new{can be block
diagonalised to give the irreducible representations of the $A_4$ group}
%
\begin{equation*}
 S_{\mathrm{block\ diagonal}}=\begin{pmatrix}
              1&\cdots & 0 &\cdots \\
              \vdots &\ddots & & \\
              0 & & S_3 & \\
              \vdots& & & \ddots
             \end{pmatrix},
 T_{\mathrm{block\ diagonal}}=\begin{pmatrix}
              1&\cdots & 0 &\cdots \\
              \vdots &\ddots & & \\
              0 & & T_3 & \\
              \vdots& & & \ddots
             \end{pmatrix}
\end{equation*}
where $T_3$ and $S_3$ are the generators of $A_4$ in the 3D
irreducible representation given by:
\begin{equation}
S_3=\begin{pmatrix}
 1 & 0 & 0 \\
 0 & -1 & 0 \\
 0 & 0 & -1
\end{pmatrix},T_3=\begin{pmatrix}
                  0 & 0 & 1 \\
                  1 & 0 & 0 \\
                  0 & 1 & 0
                   \end{pmatrix}.
\end{equation}

\subsection{Parametrising multiplets}

If we are to place fields at the fixed points of the orbifold then
we will need to parametrise a 4 dimensional representation in
terms of singlet and triplet representations as in
\cite{Altarelli:2006kg}. We now briefly \new{summarise the results of}
\cite{Altarelli:2006kg} to build the dictionary in Table \ref{dictionary} from a 6d
orbifolded theory to an effective 4d one. If we consider a
multiplet $u=(u_1,u_2,u_3,u_4)^T$ transforming as:
\begin{eqnarray*}
 \mathcal{S}:u&\rightarrow &Su \\
\mathcal{T}:u&\rightarrow &Tu,
\end{eqnarray*}
We can decompose the
reducible quadruplet into a triplet and invariant singlet
irreducible representations:
\begin{equation*}
  \begin{pmatrix}
  u_1 \\
  u_2 \\
  u_3 \\
  u_4
 \end{pmatrix}=\frac{1}{2}\begin{pmatrix}
                           v_0 \\
                           v_0 \\
                           v_0 \\
                           v_0
                          \end{pmatrix}+\frac{1}{2}\begin{pmatrix}
                                                     -v_1+v_2+v_3 \\
                                                     +v_1-v_2+v_3 \\
                                                     +v_1+v_2-v_3 \\
                                                     -v_1-v_2-v_3
                                                    \end{pmatrix}
\end{equation*}


As noted in \cite{Altarelli:2006kg} this parametrisation is not
unique,
Brane singlets are given by a vector
of the form $a_{singlet}=(a_c/2,a_c/2,a_c/2,a_c/2)^T$, i.e. brane
fields having the same value at each fixed point. Brane Triplets
$a=(a_1,a_2,a_3)$are in one of three representations
$\mathcal{R}_{0,\pm 1}$ given by
\begin{equation}
\label{definedreps}
a^{\mathcal{R}_1}=a^{\mathcal{R}_{-1}*}=\frac{1}{2}\begin{pmatrix}
                              -a_1+\omega a_2+\omega^2a_3\\
                              +a_1-\omega a_2+\omega^2a_3 \\
                              +a_1+\omega a_2-\omega^2a_3 \\
                              -a_1-\omega a_2-\omega^2a_3
                             \end{pmatrix}, \ \
 a^{\mathcal{R}_0}=\frac{1}{2}\begin{pmatrix}
                         -a_1+a_2+a_3 \\
                         +a_1-a_2+a_3 \\
                         +a_1+a_2-a_3 \\
                         -a_1-a_2-a_3
                         \end{pmatrix}
\end{equation}
depending on which singlet the triplets are forming in the
superpotential.
Bulk singlets depend on the extra coordinates and transform as $S\mathbf{\xi}(z)=\mathbf{\xi}(z+1/2)$ and $ T\mathbf{\xi}(z)=\mathbf{\xi}(\omega z)$.
We require these decompositions because we will want to construct non-invariant singlets from
products of triplets and if we were to restrict ourselves to the
first parametrisation we would be unable to do so.

\subsection{Bulk and Brane Fields}

Following \cite{Altarelli:2006kg} we now look at the coupling of a
bulk multiplet:
$\mathbf{B}(z)=(\mathbf{B}_1(z),\mathbf{B}_2(z),\mathbf{B}_3(z)$,
transforming as a triplet of $A_4$ and the brane triplet
$a=(a_1,a_2,a_3,a_4)$ transforming as $\mathcal{R}_0$, as in
Eqn(\ref{definedreps}). The transformations of $\mathbf{B}$ are:
\begin{eqnarray*}
 \mathcal{S}:\mathbf{B}'(z_S)=S_3\mathbf{B}(z)& & z_S=z+\frac{1}{2} \\
 \mathcal{S}:\mathbf{B}'(z_T)=T_3\mathbf{B}(z)& & z_T=\omega z.
\end{eqnarray*}
We can write a bilinear in $a$ and $\mathbf{B}$ given by:
\begin{equation}
 J=\sum_{iK}\alpha_{iK}a_i^{\mathcal{R}_0}\mathbf{B}_{K}(z)\delta_i
\end{equation}
where $\alpha_{iK}$ is a four by three matrix of constant
coefficients, and $\delta_i=\delta(z-z_i)$ where $z_i$ are the fixed points. We want J to be invariant under $A_4$  then we
choose:
\begin{equation*}
 \alpha_{iK}=\frac{1}{2}\begin{pmatrix}
                         -1&+1&+1 \\
                         +1&-1&+1 \\
                         +1 &+1&-1 \\
                         -1&-1&-1
                        \end{pmatrix}.
\end{equation*}
Since $a$ is in the $\mathcal{R}_0$ representation after
integration and \new{if the triplet $\mathbf{B}(z)$ aquires a constant VEV
$\braket{\mathbf{B}(z)}=(\mathbf{B}_1,\mathbf{B}_2,\mathbf{B}_3)$
then J becomes:}
\begin{equation*}
 J=v_1\mathbf{B}_1+v_2\mathbf{B}_2+v_3\mathbf{B}_3.
\end{equation*}
We can do the same for a bilinear $J'$ given by:
\begin{equation*}
 J'=\sum_{iK}\alpha_{iK}'a_i\mathbf{B}_K(z)\delta_i
\end{equation*}
which transforms as a $1'$ with the matrix $\alpha_{iK}'$ given
by:
\begin{equation*}
 \alpha_{iK}'=\frac{1}{2}\begin{pmatrix}
                          -1&+\omega&+\omega^2 \\
                          +1&-\omega&+\omega^2 \\
                          +1&+\omega&-\omega^2 \\
                          -1&-\omega&-\omega^2
                         \end{pmatrix}.
\end{equation*}
After integrating over $z$ and after $\mathbf{B}$ has aquired a
constant VEV we find that:
\begin{equation*}
 J'=v_1\mathbf{B}_1+\omega
 v_2\mathbf{B}_2+\omega^2v_3\mathbf{B}_3.
\end{equation*}
We can obtain the $1''$ singlet by simply substituting
$\alpha_{iK}'$ by its complex conjugate to get $\alpha_{iK}''$.

\end{document}